\documentclass[twocolumn,aps,prl]{revtex4}
\usepackage{graphicx}

\begin{document}
\bibliographystyle{prsty}

\title{Probing the superconducting gap symmetry of PrOs$_{4}$Sb$_{12}$: A
penetration depth study}
\author{Elbert E. M. Chia}
\author{M. B. Salamon}
\affiliation{Department of Physics, University of Illinois at Urbana-Champaign, 1110 W.
Green St., Urbana IL 61801}
\author{H. Sugawara}
\author{H. Sato}
\affiliation{Department of Physics, Tokyo Metropolitan University, Hachioji, Tokyo
192-0397, Japan}
\date{\today}

\begin{abstract}
We report measurements of the magnetic penetration depth $\lambda$ in single
crystals of PrOs$_{4}$Sb$_{12}$ down to $\sim$0.1~K using a tunnel-diode
based, self-inductive technique at 21 MHz, with the ac field applied along
the \textit{a}, \textit{b} and \textit{c} directions. In all three field
orientations the penetration depth and superfluid density $\rho_{s}$ tend to
follow a power law, $\lambda \sim \mathit{T}^{2}, \rho_{s} \sim \mathit{T}%
^{2}$ at low temperatures. $\rho_{s}$ for various gap functions
have been calculated, and data is best fit by the $^{3}$He
A-phase-like gap with multidomains, each having two point nodes
along a cube axis, and parameter
$\Delta_{0}$(0)/\textit{k}$_{B}$\textit{T}$_{c}$ = 2.6. This
suggests that PrOs$_{4}$Sb$_{12}$ is a strong-coupling
superconductor with two point nodes on the Fermi surface. We also
confirm the observation of the double transitions at 1.75~K and
1.85~K seen in other measurements.
\end{abstract}

\maketitle

The recent discovery of the Heavy Fermion (HF) skutterudite superconductor
(SC) PrOs$_{4}$Sb$_{12}$ \cite{Bauer02,Maple02} has attracted much interest
due to its differences with the other unconventional SC, and in particular,
the HFSC. PrOs$_{4}$Sb$_{12}$ has a nonmagnetic ground state of localized
\textit{f} electrons in the crystalline electric field, hence its HF
behavior, and consequently the origin of its superconductivity, might be
attributed to the interaction between the electric quadrupolar moments of Pr$%
^{3+}$ and the conduction electrons. It is thus a candidate for the first SC
mediated by quadrupolar fluctuations, i.e. by neither electron-phonon nor,
as with other HFSC, magnetic interactions.

Recent experiments on PrOs$_{4}$Sb$_{12}$ give conflicting evidence to the
nature of the SC gap. Muon-spin rotation ($\mu$SR) measurements revealed a
low-temperature exponential behavior, suggesting isotropic pairing (either
\textit{s} or \textit{p}-wave) \cite{MacLaughlin02}. Scanning tunneling
spectroscopy measurements also measured a density of states (DOS) with no
low-energy excitations with a well-developed SC gap over a large part of the
Fermi Surface (FS) \cite{Suderow03}. The absence of a Hebel-Slichter peak
and the non-T$^{3}$ behavior of 1/T$_{1}$ in nuclear quadrupolar resonance
(NQR) experiments suggest that PrOs$_{4}$Sb$_{12}$ has a full gap or point
nodes, but not line nodes, at zero field \cite{Kotegawa03,Ichioka03}. If PrOs%
$_{4}$Sb$_{12}$ has an isotropic gap, then it is unique among HFSC,
suggesting the possibility of (a) an important difference in superconducting
properties between HFSC with magnetic and non-magnetic \textit{f}-ion ground
states, and (b) a correlation between pairing symmetry (isotropic or nodal
gap) and mechanism (quadrupolar or magnetic fluctuations) of
superconductivity \cite{MacLaughlin02}.

Unlike the $\mu$SR and NQR results, angle-dependent thermal conductivity
measurements \cite{Izawa03} revealed two distinct SC phases with different
symmetries, a phase transition between them, and presence of \textit{point}
nodes. In the high-field phase four point nodes ([100] and [010] directions)
have been observed, whereas there are only two point nodes ([010] directions
only) in the low-field phase. Specific-heat data \cite{Bauer02} also show a
low-temperature power law behavior, suggesting the presence of nodes.
Another recent $\mu$SR experiment revealed the spontaneous appearance of
static internal magnetic fields below \textit{T}$_{c}$, providing evidence
that the SC state is a time-reversal-symmetry-breaking (TRSB) state \cite%
{Aoki03}.

In this paper, we present high-precision measurements of penetration depths $%
\lambda $(\textit{T}) of PrOs$_{4}$Sb$_{12}$ at temperatures down to 0.1~K.
The ac field was applied along all three crystallographic axes. In all three
field orientations both $\lambda $(\textit{T}) and superfluid density $%
\rho_{s}$(\textit{T}) tends to follow a quadratic power law, suggesting that
the SC gap has nodes on the FS. $\rho_{s}$ for various gap functions has
been calculated, and data are best fit by the He-3 A-phase-like gap, with
two point nodes in the [010] directions. Our data thus puts PrOs$_{4}$Sb$%
_{12}$ in line with other HFSC, in that they all have nodes on the FS,
despite the proposed non-magnetic nature of the mechanism of its
superconductivity.

Details of sample growth and characterization are described in Ref.~%
\onlinecite{Sugawara02}. Measurements were performed utilizing a 21-MHz
tunnel diode oscillator \cite{Bonalde2000} with a noise level of 2 parts in
10$^{9}$ and low drift. The magnitude of the ac field was estimated to be
less than 5 mOe. The cryostat was surrounded by a bilayer Mumetal shield
that reduced the dc field to less than 1 mOe. The sample was aligned inside
the probing coil in all three crystallographic directions. The sample was
mounted, using a small amount of GE varnish, on a single crystal sapphire
rod. The other end of the rod was thermally connected to the mixing chamber
of an Oxford Kelvinox 25 dilution refrigerator. The sample temperature is
monitored using a calibrated RuO$_{2}$ resistor at low temperatures (\textit{%
T}$_{base}$ - 1.8~K), and a calibrated Cernox thermometer at higher
temperatures (1.3~K - 2.5~K).

The deviation $\Delta \lambda_{i} $(\textit{T}) = $\lambda_{i} $(\textit{T})
-- $\lambda_{i} $(0.1~K) (\textit{i} = \textit{a, b, c}) is proportional to
the change in resonant frequency $\Delta $\textit{f}(\textit{T}), with the
proportionality factor \textit{G} dependent on sample and coil geometries.
The subscript \textit{i} denotes the direction of the applied magnetic
field. For our \textit{H}//\textit{a} data, we determine \textit{G} from a
single-crystal sample of pure Al by fitting the Al data to extreme nonlocal
expressions and then adjusting for relative sample dimensions \cite{Chia03}.
Testing this approach on a single crystal of Pb, we found good agreement
with conventional BCS expressions. The value of \textit{G}$_{a}$ obtained
this way has an error of $\pm$10~\%, since our sample has a rectangular
basal area instead of square \cite{Prozorov2000}. To obtain \textit{G}$_{b}$
and \textit{G}$_{c}$ we make use of the cubic symmetry of the crystal and
assume that the total change in penetration depth from the three
orientations are equal, i.e. $\Delta \lambda_{a}$(\textit{T}$_{c}$) = $%
\Delta \lambda_{b}$(\textit{T}$_{c}$) = $\Delta \lambda_{c}$(\textit{T}$_{c}$%
). From this equality, and the value of \textit{G}$_{a}$, we can calculate
\textit{G}$_{b}$ and \textit{G}$_{c}$.

\begin{figure}
\centering \includegraphics[width=8cm,clip]{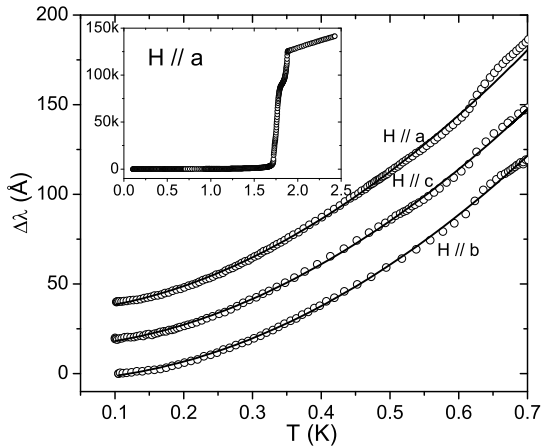}
\caption{Low-temperature dependence of the penetration depth
$\Delta \lambda$(\textit{T}) for field orientations
\textit{H}//\textit{a, b, c}. The curves have been offset for
clarity. The solid lines are fits to $\Delta \lambda$(\textit{T})
= \textit{A} + \textit{BT}$^{n}$ from 0.1~K to 0.55~K. Upper inset
shows $\Delta \lambda_{a}$(\textit{T}) over the full temperature
range.} \label{fig:lambda}
\end{figure}

Fig.~\ref{fig:lambda} shows $\Delta \lambda_{i} $(\textit{T}) as functions
of temperature. All three curves vary strongly at low temperatures,
inconsistent with exponential behavior expected for isotropic \textit{s}%
-wave superconductors. On the other hand, the variation is not linear, but
has an obvious upward curvature, unlike the low-temperature behavior
expected for pure \textit{d}-wave superconductors. A fit of the low
temperature data (up to 0.55~K $\approx$ 0.3~\textit{T}$_{c}$) to a variable
power law, $\Delta \lambda$(T) = \textit{A} + \textit{BT}$^{n}$ yields
\textit{n} = 1.9 $\pm $ 0.1 for \textit{H}//\textit{a}, \textit{b}, and
\textit{n} = 2.0 $\pm $ 0.1 for \textit{H}//\textit{c}. The uncertainty in
\textit{n} is \textit{not} a consequence of the uncertainty in \textit{G},
but rather of the somewhat uneven faces of the crystal and the range of fit.
Within the uncertainty in \textit{G} the three curves are consistent with
one another. There is also a small upturn near 0.62~K, which might distort
the low-temperature fit and cause the power law to deviate from \textit{T}$%
^{2}$. The NQR spin-lattice relaxation rate also changes around
this temperature, however, the origin is not clear at present
\cite{Kotegawa03,Sugawaraemail}. A non-unitary state has the
unique feature that spin-up and spin-down Cooper pairs have
different excitation gaps \cite{Sigrist91}. If the SC state in
PrOs$_{4}$Sb$_{12}$ is a TRSB state, then this upturn may be due
to the contribution from the smaller gap \cite{Aoki03}. It is
interesting to note that a fit of the $\Delta \lambda
$(\textit{T}) from 0.6~K to 1.1~K, to the same variable power law,
gives an exponent of about 3, consistent with \textit{n} $\approx
$ 4 obtained for specific-heat data over the same temperature
range \cite{Bauer02}.

\begin{figure}
\centering \includegraphics[width=8cm,clip]{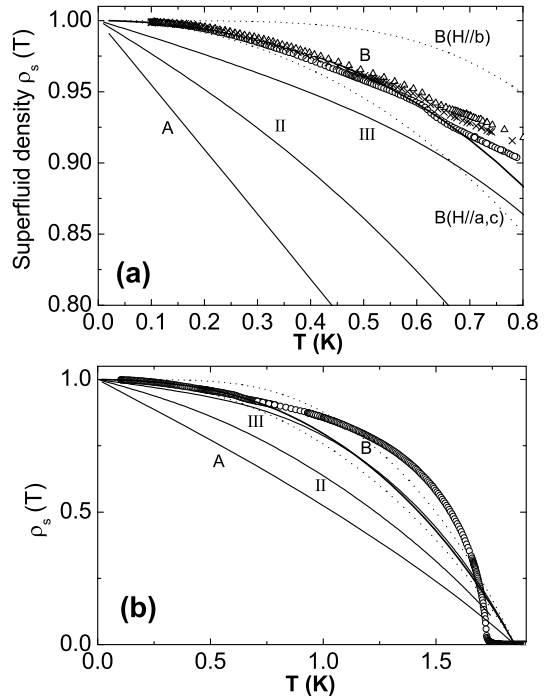} \caption{(a)
Low-temperature superfluid density $\rho_{s}$(\textit{T}) =
[$\lambda^{2}$(0)/$\lambda^{2}$(\textit{T})] calculated from
$\Delta \lambda $(\textit{T}) data in Fig.~\ref{fig:lambda}, for
all three field orientations. ($\bigcirc$) \textit{H}//
\textit{a}, ($\bigtriangleup$) \textit{H}//\textit{b}, ($\times$)
\textit{H}//\textit{c}. Using
$\Delta_{0}$(0)/\textit{k}$_{B}$\textit{T}$_{c}$ = 2.6, the solid
lines are the calculated effective superfluid density
$\rho_{s}^{eff}$ corresponding to gaps II, III, A and B. The
dotted lines correspond to $\rho_{s}$($H$//$b$) and
$\rho_{s}$($H$//$a$, $c$) for gap B. (b) The same calculated
curves over the entire temperature range. Only the
$\rho_{s}$(\textit{T}) data for $H$//$a$ are shown.}
\label{fig:rho}
\end{figure}

Using the value of $\lambda$(0) = 3440 \AA\ from $\mu$SR measurements \cite%
{MacLaughlin02}, we calculated the superfluid density $\rho_{s} $ from our
data. We follow the procedure in Ref.~\onlinecite{Chia03} to compute the
experimental superfluid density, using the \textit{T}$^{2}$ fit to estimate
the small difference between $\lambda$(0) and $\lambda$(0.1~K). Fig.~\ref%
{fig:rho}a shows $\rho_{s} $(\textit{T}) for all three field orientations at
low temperatures. In each case, a fit of $\rho_{s} $(\textit{T}) to a
variable power law, $\rho_{s} $(\textit{T}) = 1 - $\alpha$\textit{T}$^{n}$
also yields n $\approx$ 2, from 0.1~K ($\sim$ 0.05~\textit{T}$_{c}$) to
0.55~K. Once again this suggests the presence of low-lying excitations,
incompatible with an isotropic SC gap.

\begin{figure}
\centering \includegraphics[width=8cm,clip]{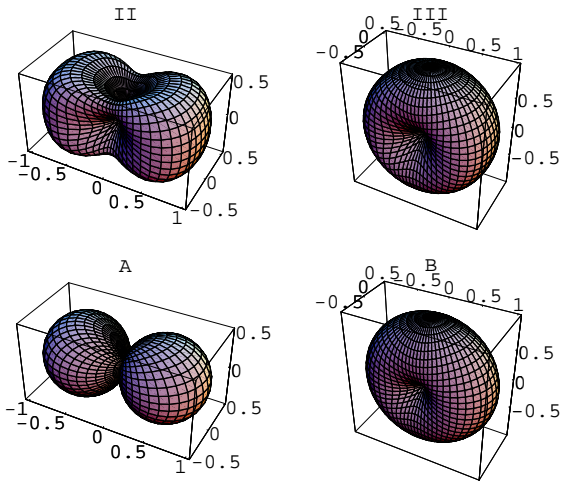} \caption{Polar
plots of gap functions (II), (III), (A) and (B).}
\label{fig:Polarplot}
\end{figure}

Several theoretical proposals have been put forward to understand
the two SC phases \cite{Maki02,Goryo03,Ichioka03}. To explain the
behavior of the angle-dependent, magneto-thermal-conductivity
results \cite{Izawa03}, Maki \textit{et al.} \cite{Maki02}
proposed three possible SC gap functions for PrOs$_{4}$Sb$_{12}$.
In particular, for the low-field (L) phase, two gap
functions were proposed: (II) \textit{f}($\mathbf{k}$)=1 - \textit{k}$%
_{y}^{4}$ - \textit{k}$_{z}^{4}$ having four nodes and (III) \textit{f}($%
\mathbf{k}$)=1 - \textit{k}$_{y}^{4},$ with two point nodes. The gap
function is $\Delta $($\mathbf{k}$) $\equiv \Delta _{0}$\textit{f}($\mathbf{k%
}$), with the form factor \textit{f}($\mathbf{k}$) normalized to unity and $%
\Delta _{0}$ the temperature-\textit{dependent} maximum gap value.
As we will see, both functions lead to a linear temperature
dependence for the superfluid density. \ Consequently, we consider
two further gap functions: (A)
\textbf{\textit{d}}(\textbf{k})=$\hat{y}$ \textit{k}$_{a}$ (a
line-node gap), and (B)
\textbf{\textit{d}}(\textbf{k})=$\hat{y}$ (\textit{k}$_{a}\pm $ i\textit{k}$%
_{c}$), for which $\Delta $(\textbf{k})= $\Delta _{0}$ $\mid $\textbf{%
\textit{d}}(\textbf{k})$\mid .$ \ Gap B has two point nodes along
the [010] directions and a gap dispersion identical to the
superfluid $^{3}$He A-phase, $\Delta $($\mathbf{k}$) = $\Delta
_{0}\left\vert k_{a}\pm ik_{b}\right\vert $ = $\Delta _{0}\sin
\theta $; hence, they give identical temperature dependences of
$\rho _{s}$. Polar plots of these gap functions are shown in
Fig.~\ref{fig:Polarplot}. We have assumed the gap maximum $\Delta
_{0}$(\textit{T}) to have the form $\Delta _{0}(\mathit{T})=\delta
_{sc}\mathit{kT}_{c}tanh\{\frac{\pi
}{\delta _{sc}}\sqrt{a(\frac{\Delta C}{C})(\frac{T_{c}}{T}-1)}\}$ \cite%
{Gross1986}, where $\delta _{sc}\equiv \Delta _{0}$(0)/\textit{k}$_{B}$%
\textit{T}$_{c}$ is the only variable parameter, \textit{T}$_{c}$
= 1.85~K, \textit{a} = 2/3, and the specific heat jump $\Delta
$\textit{C}/\textit{C}~=~3 is an experimentally obtained value
\cite{Vollmer03}.

A problem arises immediately with point nodes. If there are only
two point nodes in the [010] directions, breaking cubic symmetry,
then one would expect $\Delta \lambda _{b}$ to tend toward an
exponential temperature dependence at low temperatures. We show
this in Fig.~\ref{fig:rho}, where we have calculated the
superfluid density for gap B for fields along [010] and either
[100] or [001]. A measurement along [010] would indeed give
exponential behavior while measurements in orthogonal directions
give a strong temperature dependence. However, our experimental
data show otherwise - there is an almost identical
\textit{T}$^{2}$ superfluid response in all three field
orientations. While it is possible that, in the absence of an
external agent, the sample will randomly choose one, {\it and only
one}, axis along which to locate the nodes each time it becomes
superconducting, it is much more likely to develop a domain
structure. One possibility is that the SC order parameter and
strain are coupled \cite{Goryo03}, a situation similar to
magnetostriction in ferromagnets. Such a situation also arises in
Cromium (Cr) \cite{Fawcett88}, where the coupling between the
spin-density-wave (SDW) and strain wave causes the wave vector
\textbf{Q} of the modulation to point along any \{001\} direction
in the bcc Cr lattice. In bulk Cr all three possible orientations
occur with equal probability. A Cr single crystal thus has
multiple domains, with each domain corresponding to one of three
possible
\textbf{Q}$_{x}$, \textbf{Q}$_{y}$ and \textbf{Q}$_{z}$ regions \cite%
{Werner66} - the ``poly-\textbf{Q}" state. Tensile stress in
applied along one cube axis while cooling through
\textit{T}$_{N}$, however, produces a ``single-\textbf{Q}" state
with all domains having their \textbf{Q}'s pointing in the same
direction, along the stress axis \cite{Bastow66}.
Evidence that a domain with only two point nodes can form in PrOs$_{4}$Sb$%
_{12}$ was reported in Ref.~\onlinecite{Izawa03}, where nodes were seen only
along a single [010] axis. We suggest that the experimental setup of Ref.~%
\onlinecite{Izawa03} may have strained the sample to produce a
single domain, analogous to the single-\textbf{Q} state of Cr.

\begin{figure}
\centering \includegraphics[width=8cm,clip]{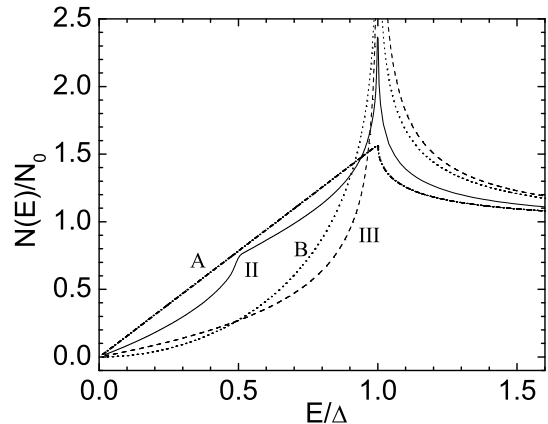}
\caption{Quasiparticle DOS for the gap functions II (full line),
III (short-dashed line), A (short-dash-dotted line) and B
(short-dotted line).} \label{fig:DOS}
\end{figure}

Assuming the existence of domains, we plot an effective superfluid
density $\rho_{s}^{eff}$ by taking the average of [100], [010] and
[001]-superfluid densities, with equal weight from each component.
The superfluid densities in different directions are evaluated
using the expression

\begin{equation}
\rho_{s} (H//x) = 1 - \frac{3}{N(0)} \sum_{\textbf{k}}
(\hat{k}^{2}_{y} + \hat{k}^{2}_{z}) \frac{\partial f}{\partial
E_{\textbf{k}}}
\end{equation} where {\it x, y, z} = any permutation
of {\it a, b, c}. $N$(0) is the quasiparticle DOS at the FS, $f$ =
[exp(E$_{\textbf{k}}$/{\it k}$_{B}T$)+1]$^{-1}$ is the Fermi
function, and $E_{\textbf{k}}$ = [$\varepsilon^{2}(\textbf{k})$ +
$\Delta(\hat{\textbf{k}},T)^{2}$]$^{1/2}$ is the quasiparticle
energy. The component superfluid densities for Gap B are shown as
two dotted lines, and $\rho_{s}^{eff}$ as a solid line, in
Fig.~\ref{fig:rho}. Clearly, the agreement between data and
$\rho_{s}^{eff}$ is very good. We chose the strong-coupling value
$\delta _{sc}$ = 2.6 here, taken from
Ref.~\onlinecite{Kotegawa03}. Using $\delta _{sc}$ = 2.1 from
Ref.~\onlinecite{MacLaughlin02} gives a worse fit. For the other
gap functions, we also calculated $\rho_{s}^{eff}$ (shown in
Fig.~\ref{fig:rho}) and $\rho_{s}$({\it H//a, b, c}) (not shown
here) - all of them give linear temperature dependences and fall
far from the experimental data. The effective quasiparticle DOS
for all four gap functions are also shown in Fig.~\ref{fig:DOS}.
Our data therefore suggest that PrOs$_{4}$Sb$_{12}$ is a
strong-coupling unconventional SC. The superfluid data is best fit
with a $^{3}$He A-phase-like gap, with two point nodes on its FS.
Both the field-direction-independence of the superfluid data,
$and$ the nice fit of the data to $\rho_{s}^{eff}$, strongly
suggest that PrOs$_{4}$Sb$_{12}$ has multidomains. Note that
though Gap B is a unitary gap, our low-$T$ data can also be fit
\cite{Machidaemail} by the two-point-node L-phase
\textit{non-unitary} gap proposed by Ichioka \textit{et al.} in
Ref.~\onlinecite{Ichioka03}, with similar DOS structure. Hence our
data does not contradict the non-unitary result of Aoki \textit{et
al.} \cite{Aoki03}.

It is already apparent in Fig. \ref{fig:rho}a that the data
deviate from the proposed gap function above 0.6~K. This is even
clearer in Fig.~\ref{fig:rho}b, which shows $\rho
_{s}$(\textit{T}) for \textit{H}//\textit{a} from 0.1~K to
\textit{T}$_{c}$. The other two field orientations give almost
identical temperature dependence. None of the four gap functions
fits the data over this larger temperature range. This could be
due to the opening up of the smaller gap caused by the
non-unitarity of the SC state, as mentioned earlier. Also, strong
changes in the mass renormalization in different sheets have been
found in de Haas van Alphen experiments \cite{Sugawara02}. These
changes may cause the distribution of values of the SC gap
measured in tunneling measurements \cite{Suderow03}, and
strengthens both the idea that the mass renormalization and
superconductivity are of the same origin, i.e. that the
quadrupolar fluctuations favor SC correlations, as well as the
possible multiband character of superconductivity in this material
\cite{Suderow03}. Hence a multiband analysis, similar to those
performed on MgB$_{2}$, might be required to fit the superfluid
data over the entire temperature range.

\begin{figure}
\centering \includegraphics[width=8cm,clip]{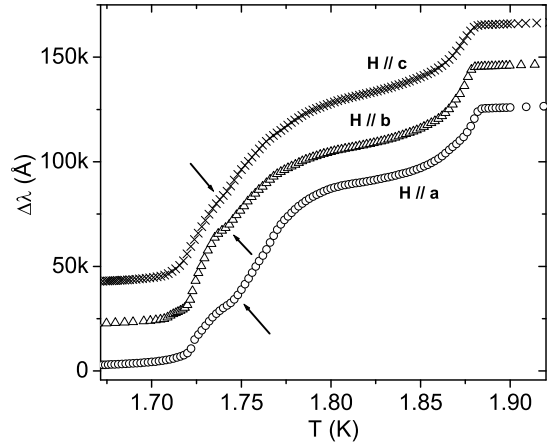} \caption{$\Delta
\lambda$(\textit{T}) for all three field orientations near
\textit{T}$_{c}$. The curves have been offset for clarity. The
arrows indicate the second superconducting transition at 1.74~K.}
\label{fig:lambdaTc}
\end{figure}

Finally, we turn to the region near \textit{T}$_{c}$. Fig.~\ref{fig:lambdaTc}
shows $\Delta \lambda $(\textit{T}) near \textit{T}$_{c}$ for all three
field orientations. Three features can be seen: the onset of
superconductivity at 1.88~K, a strong but broad shoulder near 1.8~K, and
finally a weak shoulder near 1.74~K (observable even in the \textit{H}//%
\textit{c} data). In another sample from the same batch, only the first and
third features were observed. The 1.88~K and 1.74~K features confirm the two
superconducting transitions seen in the specific-heat measurement \cite%
{Vollmer03}, and suggested by angle-dependent thermal conductivity
measurements \cite{Izawa03}. The origin of the 1.8~K shoulder is unknown. In
the $\rho _{s}$(\textit{T}) plot, $\rho _{s}$ already approaches zero near
1.7~K. So these features were not discernible there. Also, we did not see
any anomaly around \textit{T}$^{\ast }$~=~2.3~K that was observed in Ref.~%
\onlinecite{Kotegawa03}.

In conclusion, we report measurements of the magnetic penetration depth $%
\lambda $ in single crystals of PrOs$_{4}$Sb$_{12}$ down to 0.1~K
using a tunnel-diode based, self-inductive technique at 21 MHz,
with the ac field applied along the \textit{a}, \textit{b} and
\textit{c} directions. In all three field orientations $\lambda $
and superfluid density $\rho _{s}$ tend to follow a quadratic
power law. We have calculated $\rho _{s}$ for various gap
functions, finding that the data are best fit by the $^{3}$He
A-phase-like gap function with two point nodes on the FS. We also
observe the double transitions near 1.75~K and 1.85~K seen in
other measurements.

One of the authors (E.E.M.C.) acknowledges D. Lawrie and Y.
Matsuda for useful discussions. Special thanks to K. Machida for
significant contributions. This work was supported by the NSF
through Grant No. DMR99-72087.

\bibliographystyle{prsty}
\bibliography{PrOs4Sb12,CeCoIn5v11}

\begin{thebibliography}{10}

\bibitem{Bauer02}
E.~D. Bauer {\it et~al.}, Phys. Rev. B {\bf 65},  100506 (R)  (2002).

\bibitem{Maple02}
M.~B. Maple {\it et~al.}, J. Phys. Soc. Jpn., Suppl. B {\bf 71},  23  (2002).

\bibitem{MacLaughlin02}
D.~E. MacLaughlin {\it et~al.}, Phys. Rev. Lett. {\bf 89},  157001  (2002).

\bibitem{Suderow03}
H. Suderow {\it et~al.}, cond-mat/0306463  (2003).

\bibitem{Kotegawa03}
H. Kotegawa {\it et~al.}, Phys. Rev. Lett. {\bf 90},  027001  (2003).

\bibitem{Ichioka03}
M. Ichioka, N. Nakai, and K. Machida, J. Phys. Soc. Jpn. {\bf 72},  1322
  (2003).

\bibitem{Izawa03}
K. Izawa {\it et~al.}, Phys. Rev. Lett. {\bf 90},  117001  (2003).

\bibitem{Aoki03}
Y. Aoki {\it et~al.}, Phys. Rev. Lett. {\bf 91},  067003  (2003).

\bibitem{Sugawara02}
H. Sugawara {\it et~al.}, Phys. Rev. B {\bf 66},  220504 (R)  (2002).

\bibitem{Bonalde2000}
I. Bonalde {\it et~al.}, Phys. Rev. Lett. {\bf 85},  4775  (2000).

\bibitem{Chia03}
E.~E.~M. Chia {\it et~al.}, Phys. Rev. B {\bf 67},  014527  (2003).

\bibitem{Prozorov2000}
R. Prozorov, R.~W. Gianetta, A. Carrington, and F.~M. Araujo-Moreira, Phys.
  Rev. B {\bf 62},  115  (2000).

\bibitem{Sugawaraemail}
H. Sugawara, private conversation  .

\bibitem{Sigrist91}
M. Sigrist and K. Ueda, Rev. Mod. Phys. {\bf 63},  239  (1991).

\bibitem{Maki02}
K. Maki {\it et~al.}, cond-mat/0212090  (2002).

\bibitem{Goryo03}
J. Goryo, Phys. Rev. B {\bf 67},  184511  (2003).

\bibitem{Gross1986}
F. Gross {\it et~al.}, Z. Phys. B {\bf 64},  175  (1986).

\bibitem{Vollmer03}
R. Vollmer {\it et~al.}, Phys. Rev. Lett. {\bf 90},  057001  (2003).

\bibitem{Fawcett88}
E. Fawcett, Rev. Mod. Phys. {\bf 60},  209  (1988).

\bibitem{Werner66}
S.~A. Werner, A. Arrott, and H. Kendrick, J. Appl. Phys. {\bf 37},  1260
  (1966).

\bibitem{Bastow66}
T.~J. Bastow and R. Street, Phys. Rev. {\bf 141},  51  (1966).

\bibitem{Machidaemail}
K. Machida, private conversation  .

\end{thebibliography}
\bigskip

\end{document}